# A Novel Similarity Measure for Intrusion Detection using Gaussian Function

**Gunupudi Rajesh Kumar[1], Mangathayaru Nimmala[2], G.Narsimha[3]**

[1]Faculty of Information Technology, VNRVJIET, Hyderabad, 500090, India
[2]Professor, Deparment of Information Technology, VNRVJIET, Hyderabad, 500090, India
[3]Associate Professor, Department of CSE, JNTU, Hyderabad, India.
Email:gunupudirajesh@gmail.com[1]

**Abstract.** In this paper the major objective is to design and analyze the suitability of Gaussian similarity measure for intrusion detection. The objective is to use this as a distance measure to find the distance between any two data samples of training set such as DARPA Data Set, KDD Data Set. This major objective is to use this measure as a distance metric when applying k- means algorithm. The novelty of this approach is making use of the proposed distance function as part of k-means algorithm so as to obtain disjoint clusters. This is followed by a case study, which demonstrates the process of Intrusion Detection. The proposed similarity has fixed upper and lower bounds. The proposed similarity measure satisfies all properties of a typical similarity measure.

## 1. INTRODUCTION

The Intrusion Detection is the process of acquiring or an unauthorized attempt, to acquire the rights over computing resources or information resources. Nowadays Intrusion Detection is becoming an alarming problem. Research in this area is started many years back and there were significant improvements in the intrusion detection process. The attacks and threats are also changing their orientation while this aggression.

Several Intrusion Detection Systems are in use which are working on different approaches such as Signature based (Neminath, 2014), anomaly based, SVM Based (Thorsten, 1999), Text Processing, Genetic algorithm based, Fuzzy Logic based (Mohammad, 2011) and Association Rule based approaches. Apart from all these approaches if the intrusion detection mechanisms are to be devised in two broad categories, they are signature based and anomaly based techniques (James, 2006)(Lee, 1999).

Signature based Intrusion Detection System (IDS) is generally works based on analyzing the packets, packet sequences, traffic analysis. The IDS will search into the packet for some sequence or pattern which we call as a signature known to be malicious (Sang, 2003). The Signature based detection approach gives fruitful results only for the known attacks. The advantage of these approaches is the identifying a signature for a threat and loading its pattern into the database is quite simple (Yuxin, 2013). Once these signatures were loaded into the database, the IDS will check each packet and compare whether the signature pattern is present in the packet or not in the packet or bit sequence. The Signature matching engines do have their own disadvantages as they detect only known attacks. This approach is not suitable for those attacks which were not present in the Signature database. The rate of getting false alarms in this case is huge in size. The reason for getting false alarms very frequently is that generally a signature consists of regular expressions, string patterns. While the signature based detection shows an excellent performance in the case of threats consisting of fixed behaviour patterns. Whereas, it is almost impossible to detect unknown and threats that do frequently changes behavior as in attacks generated through intelligent softwares such as worms, Trojans, etc., as they have self- modifying behavioral characteristics. A Signature based IDS introducing the arms race between the IDS Signature developers and attackers. The performance of the signature based IDS is greatly influenced by the size of the signature data base as it has accelerated growth in volume. Even Small variation in the signature, causing a new entry in a signature database (Chirag, 2013)(Govindarajan, 2011).

The anomaly based Intrusion Detection System basically works on the principle of creating boundaries which specifies accepted behavior and unaccepted behavior. Any incoming event or outgoing event which falls in the range of unaccepted behavior in an anomaly detection engine declares it as a threat. The important point while designing the anomaly detection engine is that the engine must be given power to get into deep of each of the protocols that need to be monitored by the engine. Of course this is a very expensive job as dissection of the protocols in the initial stage is complex. The biggest challenge of the anomaly based IDS, is to understand, design, test and implement the rules for each protocol. On the other hand, if the rules are formed, the anomaly based IDS performs threat a detection job can be scaled more quickly and easily than the signature based IDS. The major pitfall of anomaly based IDS is that any anomaly within the range of the normal usage patterns is undetected. However, the anomaly based IDS is far better than the signature based IDS as any new threat not having the signature will be detected as its behavior is out of the normal behavior pattern (Alok, 2007).

In this paper, we use a novel similarity measure to form the normal behavior over the system calls caused by the processes.

## 2. VARIOUS KNOWLEDGE DISCOVERY BASED APPROACHES FOR INTRUSION DETECTION

Since intrusion detection mechanisms are imperfect, continuous surveillance of security compromises is becoming mandatory. This role will be taken care with the intrusion detection system (IDS), which aims at raising alarms whenever malicious activity is detected. Intrusion detection systems can be of two types, the host based intrusion detection system (HIDS) and the network based intrusion detection system (NIDS). The purpose of IDS is to continuously, monitor the incoming requests and correlate with the existing knowledge base and an alarm is raised if the anomaly is detected considering it as a threat. Unfortunately, so far, no intrusion detection system is proved to be perfect in this detection process. Many algorithms with different approaches were proposed in the literature, but none of them is proved to be ideal approach. Each new approach that is getting published is demonstrating, an improvement over the existing system only (Davis, 2011).

The intrusion detection systems (IDSs) can also be categorized based on the misuse and anomaly based. The misuse based intrusion detection systems generally rely upon the rules framed by domain experts. These rules play an important role in making the decision where the incoming request is actually a threat or not. The performance of the detection process is relying on the rules that were framed. All commercial software such as anti-virus software uses misuse based approach and generates very less number of false alarms. As the number of threats or malwares increases, then there is a direct influence over the correlation of the rules with knowledge base resulting lower performance. In this case the job of continuous analysis and update process is a laborious job. Adding fuel to fire, now a days is becoming a challenging that many tools are available, making the attacker's job easy, to create new threats or attacks. One of the drawbacks of the misuse based detection process is that these systems cannot detect the zero day or novel attacks.

The other approach is anomaly based approach, which was proposed to deal with novel attacks. This approach works based on the behavior of the processes. If the behavior of the process is deviating from the pre-determined model that is developed for the detection process, it raises an alarm. Unfortunately, anomaly based intrusion detection systems generate false positives that were triggered by novel but genuine traffic. This issue in anomaly based intrusion detection systems results in exhibition of poor performance. Even 1% of the false positives leads to severe impact over the performance of the IDS. Another problem with anomaly based intrusion detection is that it is also generating too much of false negatives. Anomaly based approach is one of the prominent research area, because of this reason (Davis, 2011).

The intrusion detection systems (IDSs), scans the traffic passively. There are intrusion prevention systems (IPS), which works in line with traffic, thus prevents the malicious traffic from entering the network. These techniques prevent malicious traffic before they enter the organization network avoiding severe damage to the computing resources.

## 3. PROPOSED APPROACH

In this current work, we have two objectives. To design the distance measure with tight bounds. For this we use the basic Gaussian function to obtain the proposed distance metric designed. The designed distance measure may also be used as metric to compute the similarity between two processes. There is a tight bound on the distance values computed using our current measure. In short, we have a minimum value and a maximum value of zero (0) and unity (1) respectively. In the earlier work, we adopted the distance metric on binary representation of process system-call representation (Gunupudi, 2015). In this paper, we adopt the frequency representation of the process-system call matrix. The objective is to show the proposed measure holds good for both binary and frequency representations with tight bounds on minimum and maximum possible distance values. We also discuss the space complexity if input representation is considered directly without dimensionality reduction. The space efficiency achieved is discussed in the section-5. Another objective is to use this distance measure as part of clustering algorithm and then perform classification and predict intrusion. Sometimes, the clustering process may not converge and diverges repeatedly. In such a case, it is better to stop the iterations after performing for predefined number of iterations and use these clusters to achieve required objective. A point to be remembered and noted here is that the accuracy of clustering is not the criteria for the required objective to be achieved. We are performing clustering only for the reason of achieving dimensionality reduction. To achieve dimensionality reduction, we choose to obtain number of clusters equal to a total count of the decision classes of dataset. However, obtaining clusters equal to a number of decision classes is also a relaxed constraint. The main objective is to obtain the distance of each process to cluster centers of clusters generated and add this distances to the nearest neighbor distance computed as discussed in sections below. In this way, we transform high dimensional process into low dimensional process.

### 3.1 Dimensionality Reduction of Training Set for Intrusion Detection

In this current work, we adopt the framework of the authors (Wei-Chao, 2015) for dimensionality reduction. For achieving dimensionality reduction, we cluster the process into the desired number of clusters say 'K'. This 'K' is usually the number of decision class labels of the input dataset. However, we can obtain any number of desired clusters, K < P where P is assumed to be the total number of processes being considered in dataset. To perform process of clustering, we adopt the designed Gaussian based distance measure to cluster processes, find the distance between each process and cluster centers of clusters generated and also to compute the nearest neighbor distances of intra processes. The novelty of our approach starts with the usage of distance measure designed.

In the present approach dimensionality reduction is achieved by clustering training set processes adopting the proposed measure. Here, we consider obtaining clusters from training set by setting the bound of number of clusters as a number of decision classes. However, in practice, we may obtain any number of clusters. In our case, the number of clusters is a relaxed constraint. We aim clustering of processes only for achieving reduced process dimensionality. In the discussions below, we use clustering method K-means with the proposed distance measure. The clustering achieved is static as it requires the entire processes to be specified well ahead before clustering. However, even for dynamically varying and incoming processes, we may perform clustering using proposed distance measure. This is because of the flexibility of the framework of Gaussian function adopted. Processed Intrusion datasets have labeled attacks, because of which we have flexibility to decide on the bound of clusters. If the bound is known before clustering, obviously better choice is k-Means (Wenke, 1998). In our discussions below, we have used k=2 and it is not a compulsion. We may choose a different value of k as clustering accuracy is not our objective here.

### 3.2 Distance Measure of K-Means

K-means traditionally uses distance measures such as the Euclidean, city block, and cosine etc. We choose to use a Gaussian based distance function to obtain the clusters using framework of k-means procedure. The measure designed has tighter bounds and may be used for computing process similarity or distances.

### 3.3 Gaussian Function

We consider the Gaussian function based distance measure to find the similarity between the data samples of the intrusion dataset. We use the same distance measure and apply k-Means algorithm to cluster the data samples. For the purpose of dimensionality reduction, we use the k-Means clustering technique to obtain clusters using the proposed distance function and then, to find the distance between each training data sample and each of the cluster centroids. This is further followed by finding the nearest neighbor for every data sample within the cluster. These two distances are summed to get a new distance value. This distance value becomes a singleton feature for each training data sample. Thus each data sample of the training set is mapped to a single feature value reducing the dimensionality to 1.

The Proposed distance function is defined as given in Equation. 5. We consider the Gaussian function based distance measure to find the similarity between processes of the intrusion dataset. We use the same distance measure and apply the k-Means algorithm to cluster the processes. For the purpose of dimensionality reduction, we use the k-Means clustering technique to obtain clusters using the proposed distance function and then find the distance between each training data sample and each of the cluster centroids. This is further followed by finding the nearest neighbor for every data sample within the cluster. These two distances are summed to get a new distance value. This distance value becomes a singleton feature for each training data sample. Thus each data sample of the training set is mapped to a single feature value reducing the dimensionality to 1.

The Proposed distance function is defined as given in Equation. 1

$$G(x, \mu, \sigma) = \begin{cases} e^{-\left(\frac{x-\mu}{\sigma}\right)^2}; & \text{one or both system calls exist} \\ 0 & ; \text{none of the system calls exist} \end{cases}$$

(1)

Where

x = system call being considered
μ = mean of the system call w.r.t data samples present in the cluster

σ= standard deviation of the system call considered w.r.t data samples of the training set.

The denominator of IDSIM is given by Equation.2 as shown below

$$H(x,\mu,\sigma) = \begin{cases} 1 & ;\ one\ or\ both\ system\ calls\ exist \\ 0 & ;\ none\ of\ the\ system\ calls\ exist \end{cases}$$

(2)

The average distance is the ratio of $G(x,\mu,\sigma)$ and $H(x,\mu,\sigma)$ and is represented as given by Equation 3.

$$\frac{G(x,\mu,\sigma)}{H(x,\mu,\sigma)}$$

(3)

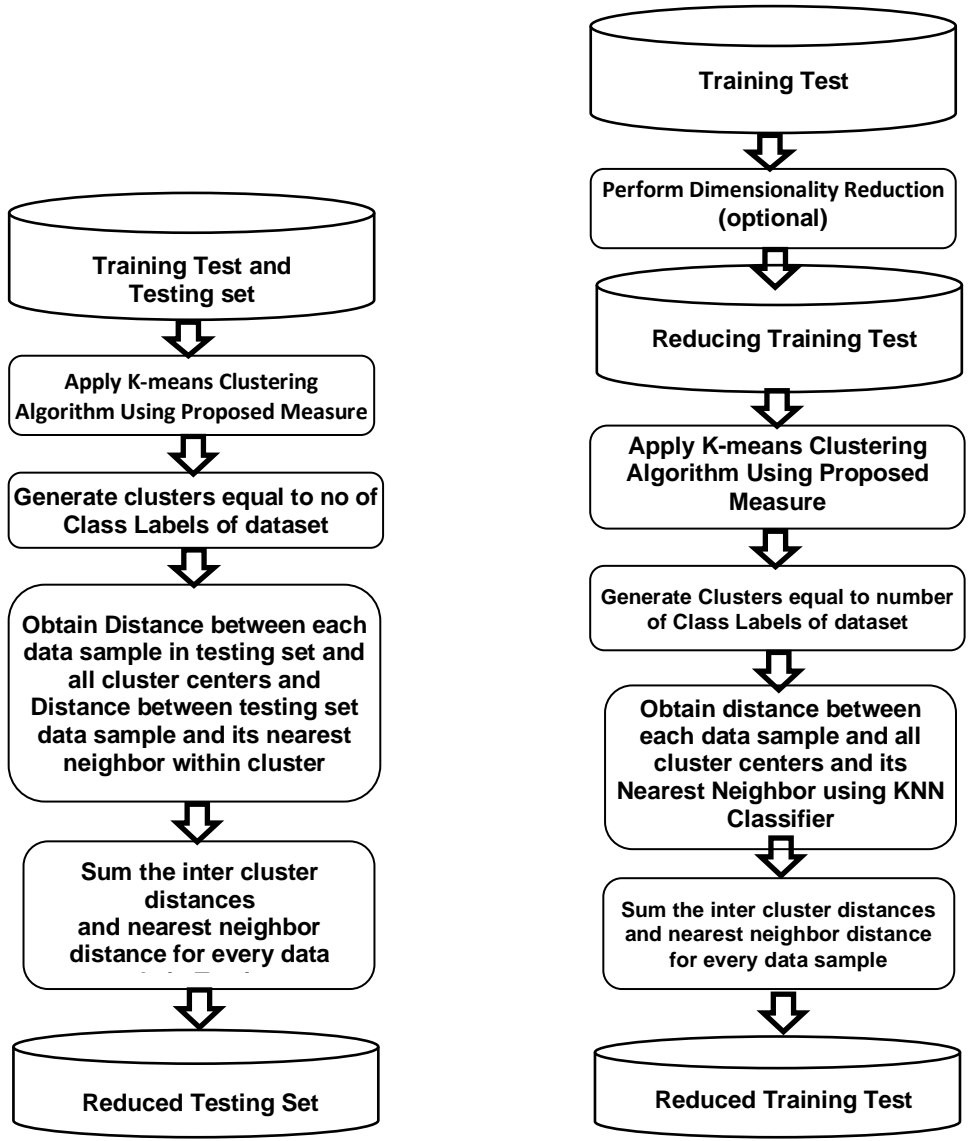

Figure 1: (a) Dimensionality Reduction of Testing Set  (b) Dimensionality Reduction of Training Test

The average distance considering the distribution of all features hence is defined as the ration of $G(x,\mu,\sigma)$ and $H(x,\mu,\sigma)$ which is reduced to Equation.4 as given below

$$F_{avg} = \frac{\sum_{i=1}^{i=n}\sum_{s=1}^{s=m} e^{-(\frac{x_{is}-\mu_{is}}{\sigma_s})^2}}{\sum_{s=1}^{s=m} 1}$$

(4)

The distance function is represented as given by

$$IDSIM = \frac{1+F_{avg}}{2}$$

(5)

Where i indicates the ith data sample. S indicates the system call. IDSIM indicates the similarity function.

$$dist = 1 - IDSIM$$

(6)

### 3.4 Dimensionality Reduction of Training Set for Intrusion Detection

Figure.1 (a) shows the proposed approach for reducing the dimensionality of the training set and Figure.1 (b) shows the proposed approach for reducing the dimensionality of the testing set using the proposed measure with k-Means clustering technique. So, we have both the testing and training sets with each data sample transformed to a singleton feature value. The test dataset can now be compared with training dataset in a very simple and effective, efficient way. The Proposed approach concentrates on using the Gaussian function based distance along with the k-Means instead of conventional distance function used by k-Means algorithm.

### 3.4 Experiments with DARPA dataset

The widely used dataset for host based intrusion detection is DARPA 98 BSM (Basic Security Module) data. The other reason for choosing DARPA is that the works by (Rawat, 2006) (Alok, 2007) were done on the DARPA 98 BSM data. Hence we prefer to choose the same dataset for our experimentation. As shown in figure 1 the DARPA dataset have 60 unique system calls and on removing 10 system calls having similar operation with small changes in prefixes and suffixes, we have obtained 50 unique system calls. Figure2 shows the list of 50final unique systems calls considered by (Rawat, 2006) (Alok, 2007), that present in training, test, and attack processes data. Hence these 50 system calls were taken into consideration for our experimentation.

Table 1  Unique System calls data in DARPA 98 BSM dataset

| List of 60 System Calls |
|---|
| access, audit, auditon, chdir, chmod, chown, close, creat, execve, exit, fchdir, fchown, fcntl, fork, fork1, getaudit, getmsg, ioctl, kill, link, login, logout, lstat, memcntl, mkdir, mmap, munmap, oldnice, oldsetgid, oldsetuid, oldutime, open, pathdonf, pipe, putmsg, readlink, rename, rmdir, setaudit, setegid, seteuid, setgroups, setpgrp, setrlimit, stat, statvfs, su, sysinfo, unlink, vfork, |
| *excluded list of system calls* |
| audition, setuid, setgid, nice, utime, ace, crew, boom, dog, eon |

Table 2 Final unique system calls data in DARPA 98 BSM dataset

| Final List of 50 System Calls |
|---|
| access, audit, auditon, chdir, chmod, chown, close, creat, execve, exit, fchdir, fchown, fcntl, fork, fork1, getaudit, getmsg, ioctl, kill, link, login, logout, lstat, memcntl, mkdir, mmap, munmap, oldnice, oldsetgid, oldsetuid, oldutime, open, pathdonf, pipe, putmsg, readlink, rename, rmdir, setaudit, setegid, seteuid, setgroups, setpgrp, setrlimit, stat, statvfs, su, sysinfo, unlink, vfork |

### 4. CASE STUDY

The processed form of DARPA Dataset (Rawat, 2006)is divided into three components, the train data, test data and attack data. The train data is consisting of 606 processes in it and all of them are normal processes. The test data comprises of 5285 normal processes enables us to test our approach. The attack data comprises of 55 attacks, each attack is having several processes in which at least one or more processes are abnormal processes. We consider only 54 attacks as one attack is exactly matching the train data.

In order to proceed for the demonstration of the proposed similarity measure, we have selected 4 normal processes randomly from the training dataset and 6 abnormal processes from the attack data from which only one process found in respective attack. It is already known that, the attack data will contain at least one or more abnormal processes; hence we select those processes which is the only process present in the attack data.

As this is only demonstration we have selected 10 system calls from the 50 system calls that present in each process. The selected 10 system calls are shown in Table 3.

**Table 3** Unique System calls data in DARPA 98 BSM dataset

| List of 10 System Calls |
|---|
| fchdir  login  pipe  logout  munmap  sysinfo audition chdir  pathdonf  boom |

The table.4 shows the process system call matrix and the corresponding class label for each process. The last column of table.4 corresponds to the class label. Here, we have two classes called normal and abnormal (Wei-Chao, 2015) (Yihua, 2002)(Rawat, 2006).

So, we choose to cluster these records in to two clusters, cluster-0 and cluster-1. For this, we use k-Means clustering algorithm as we can specify the number of clusters required by specifying the value of k=2.

But, the difference lies in the distance measure used. Here, we use the Gaussian based similarity measure for clustering using k-Means as against to traditional distance measures used such as Euclidean, Cosine, City block.

At the end of the clustering process, we have two clusters as shown in Table.5 We perform 3 iterations using k-Means by recording the clusters at each iteration. We terminate the process of clustering, when the clusters formed for two successive stages remain same. This process is shown for each iteration using the Table. 4, through Table.9

Table.13 shows the nearest neighbors for each process within the same cluster and also the similarity w.r.t clusters formed. Table.14 and table.15 shows the nearest neighbor for each process in cluster-1 and cluster-2 respectively.

**Table 4** Process system call matrix

|    | S1 | S2 | S3 | S4 | S5 | S6 | S7 | S8 | S9 | S10 | Class |
|----|----|----|----|----|----|----|----|----|----|-----|-------|
| P0 | 1  | 1  | 0  | 1  | 1  | 1  | 1  | 0  | 0  | 1   | Normal |
| P1 | 0  | 0  | 1  | 1  | 1  | 6  | 0  | 0  | 0  | 0   | Normal |
| P2 | 0  | 0  | 0  | 1  | 1  | 0  | 0  | 0  | 0  | 0   | Normal |
| P3 | 0  | 0  | 0  | 4  | 1  | 0  | 0  | 0  | 0  | 0   | Normal |
| P4 | 1  | 2  | 0  | 1  | 0  | 0  | 0  | 0  | 0  | 1   | Abnormal |
| P5 | 2  | 2  | 1  | 1  | 1  | 0  | 0  | 1  | 1  | 1   | Abnormal |
| P6 | 2  | 2  | 4  | 1  | 1  | 0  | 0  | 3  | 1  | 1   | Abnormal |
| P7 | 2  | 2  | 2  | 1  | 1  | 0  | 0  | 0  | 1  | 1   | Abnormal |
| P8 | 2  | 2  | 2  | 1  | 1  | 0  | 0  | 0  | 1  | 1   | Abnormal |
| P9 | 1  | 1  | 0  | 1  | 1  | 0  | 1  | 0  | 0  | 1   | Abnormal |

**Table 5** Initial Clusters

|    | S1 | S2 | S3 | S4 | S5 | S6 | S7 | S8 | S9 | S10 | Class |
|----|----|----|----|----|----|----|----|----|----|-----|-------|
| P0 | 1  | 1  | 0  | 1  | 1  | 1  | 1  | 0  | 0  | 1   | Normal |
| P1 | 0  | 0  | 1  | 1  | 1  | 6  | 0  | 0  | 0  | 0   | Normal |

**Table 6** Similarity of Process with Initial Clusters

|    | Cluster1 | Cluster2 | Class |
|----|----------|----------|-------|
| P0 | 1.0000   | 0.6307   | 0     |
| P1 | 0.6307   | 1.0000   | 1     |
| P2 | 0.6494 | 0.7523 | 1 |
| P3 | 0.5780 | 0.6273 | 1 |
| P4 | 0.7195 | 0.5767 | 0 |
| P5 | 0.6564 | 0.6697 | 1 |
| P6 | 0.6546 | 0.6131 | 0 |
| P7 | 0.6718 | 0.6284 | 0 |
| P8 | 0.6718 | 0.6284 | 0 |
| P9 | 0.9299 | 0.6307 | 0 |

**Table 6** Clusters

| | Processes |
|---|---|
| Cluster1 | [0, 4, 6, 7, 8, 9] |
| Cluster2 | [1, 2, 3, 5] |

**Table 7** Similarity of Process with Initial Clusters

| | Cluster1 | Cluster2 | Class |
|---|---|---|---|
| P0 | 0.7200 | 0.7129 | 0 |
| P1 | 0.6966 | 0.7525 | 1 |
| P2 | 0.7093 | 0.7525 | 1 |
| P3 | 0.6593 | 0.7467 | 1 |
| P4 | 0.7672 | 0.6402 | 0 |
| P5 | 0.8408 | 0.5994 | 0 |
| P6 | 0.7904 | 0.5731 | 0 |
| P7 | 0.8172 | 0.6164 | 0 |
| P8 | 0.8172 | 0.6164 | 0 |
| P9 | 0.7616 | 0.6945 | 0 |

**Table 8** Clusters: STAGE-2

| | Processes |
|---|---|
| Cluster1 | [0, 4, 5, 6, 7, 8, 9] |
| Cluster2 | [1, 2, 3] |

**Table 9** Similarity of Process with Initial Clusters: STAGE-3

| | Cluster1 | Cluster2 | Class |
|---|---|---|---|
| P0 | 0.7024 | 0.6094 | 0 |
| P1 | 0.6962 | 0.6484 | 0 |
| P2 | 0.7063 | 0.7074 | 1 |
| P3 | 0.6563 | 0.7051 | 1 |
| P4 | 0.7616 | 0.5510 | 0 |
| P5 | 0.8723 | 0.5690 | 0 |
| P6 | 0.8123 | 0.5586 | 0 |
| P7 | 0.8323 | 0.5659 | 0 |
| P8 | 0.8323 | 0.5659 | 0 |
| P9 | 0.7459 | 0.6083 | 0 |

**Table 11** Clusters: STAGE-3

|  | Processes |
|---|---|
| Cluster1 | [0, 1, 4, 5, 6, 7, 8, 9] |
| Cluster2 | [2, 3] |

**Table 12** Final Clusters Formed

|  | Processes |
|---|---|
| Cluster1 | [0, 1, 4, 5, 6, 7, 8, 9] |
| Cluster2 | [2, 3] |

**Table 13** Nearest Neighbours, Cluster distances, neighbour distances w.r.t each process

|  | Similarity Value w.r.t, Clusters generated | | Cluster Allotment | Nearest Neighbor | Similarity (Process, NN) |
|---|---|---|---|---|---|
|  | Cluster1 | Cluster2 |  |  |  |
| P0 | 0.7799 | 0.5780 | 0 | P9 | 0.9299 |
| P1 | 0.7140 | 0.6273 | 0 | P5 | 0.75 |
| P2 | 0.6775 | 0.7500 | 1 | P3 | 0.75 |
| P3 | 0.6275 | 0.7500 | 1 | P2 | 0.6697 |
| P4 | 0.7244 | 0.5055 | 0 | P9 | 0.7546 |
| P5 | 0.7898 | 0.5671 | 0 | P7 | 0.8773 |
| P6 | 0.7325 | 0.5648 | 0 | P5 | 0.8750 |
| P7 | 0.7562 | 0.5741 | 0 | P5 | 0.8773 |
| P8 | 0.7562 | 0.5741 | 0 | P5 | 0.8773 |
| P9 | 0.7353 | 0.5894 | 0 | P0 | 0.9299 |

**Table 14** Nearest Neighbours for Processes in Cluster 1

|  | Similarity | | | | | | | |
|---|---|---|---|---|---|---|---|---|
|  | P0 | P1 | P4 | P5 | P6 | P7 | P8 | P9 |
| P0 | 1.0000 | 0.6307 | 0.7195 | 0.6564 | 0.6546 | 0.6718 | 0.6718 | 0.9299 |
| P1 | 0.6307 | 1.0000 | 0.5767 | 0.6697 | 0.6131 | 0.6284 | 0.6284 | 0.6307 |
| P4 | 0.7195 | 0.5767 | 1.0000 | 0.6932 | 0.6909 | 0.7182 | 0.7182 | 0.7546 |
| P5 | 0.6564 | 0.6697 | 0.6932 | 1.0000 | 0.8750 | 0.8773 | 0.8773 | 0.6728 |
| P6 | 0.6546 | 0.6131 | 0.6909 | 0.8750 | 1.0000 | 0.8750 | 0.8750 | 0.6707 |
| P7 | 0.6718 | 0.6284 | 0.7182 | 0.8773 | 0.8750 | 1.0000 | 1.0000 | 0.6921 |
| P8 | 0.6718 | 0.6284 | 0.7182 | 0.8773 | 0.8750 | 1.0000 | 1.0000 | 0.6921 |
| P9 | 0.9299 | 0.6307 | 0.7546 | 0.6728 | 0.6707 | 0.6921 | 0.6921 | 1.0000 |

**Table 15** Nearest Neighbours for Processes in Cluster-2

|  | Similarity | | Nearest Neighbor |
|---|---|---|---|
|  | P2 | P3 |  |
| P2 | 1 | 0.75 | P2 |
| P3 | 0.75 | 1 | P3 |

Table.16 shows the normalized similarity value for each process which we call here as mapping value. This mapping value is obtained to obtain the dimensionality reduction. In simple words, we transform the process with dimensionality, 10 here to dimensionality of 1. In this way, we achieve dimensionality reduction of each process and finally map each process to a single value. Table.19 shows the classification process of a new test process to verify if it is normal or abnormal.

**Table 16** Calculation of Total Similarity Value and Normalized Similarity Value

|  | Similarity Value w.r.t, Clusters generated | | Clusters Generated | Nearest Neighbor | Sim(NN) | TotalSimilarity Value | Normalized FinSim =TotalSim/3 |
| --- | --- | --- | --- | --- | --- | --- | --- |
|  | Cluster1 | Cluster2 | | | | Total Similarity=sim/3 | |
| P0 | 0.7799 | 0.5780 | 0 | P9 | 0.9299 | 2.2878 | 0.7626 |
| P1 | 0.7140 | 0.6273 | 0 | P5 | 0.75 | 2.0913 | 0.6971 |
| P2 | 0.6775 | 0.7500 | 1 | P3 | 0.75 | 2.1775 | 0.7258 |
| P3 | 0.6275 | 0.7500 | 1 | P2 | 0.6697 | 2.0472 | 0.6824 |
| P4 | 0.7244 | 0.5055 | 0 | P9 | 0.7546 | 1.9845 | 0.6615 |
| P5 | 0.7898 | 0.5671 | 0 | P7 | 0.8773 | 2.2342 | 0.7447 |
| P6 | 0.7325 | 0.5648 | 0 | P5 | 0.8750 | 2.1723 | 0.7241 |
| P7 | 0.7562 | 0.5741 | 0 | P5 | 0.8773 | 2.2076 | 0.7359 |
| P8 | 0.7562 | 0.5741 | 0 | P5 | 0.8773 | 2.2076 | 0.7359 |
| P9 | 0.7353 | 0.5894 | 0 | P0 | 0.9299 | 2.2546 | 0.7515 |

**Table 17** Processes after Dimensionality Reduction using proposed measure

|  | Similarity Values | Distance |
| --- | --- | --- |
| P0 | 0.7626 | 0.2374 |
| P1 | 0.6971 | 0.3029 |
| P2 | 0.7258 | 0.2742 |
| P3 | 0.6824 | 0.3176 |
| P4 | 0.6615 | 0.3385 |
| P5 | 0.7447 | 0.2553 |
| P6 | 0.7241 | 0.2759 |
| P7 | 0.7359 | 0.2641 |
| P8 | 0.7359 | 0.2641 |
| P9 | 0.7515 | 0.2485 |

**Table 18** New test process with Nearest Neighbour, Similarity and Distance Values

|  | S1 | S2 | S3 | S4 | S5 | S6 | S7 | S8 | S9 | S10 | NearestNN | Sim | Dist |
| --- | --- | --- | --- | --- | --- | --- | --- | --- | --- | --- | --- | --- | --- |
| $P_{test}$ | 0 | 0 | 0 | 4 | 1 | 0 | 0 | 0 | 0 | 0 | Process-3 | 1.0 | 0 |
| $P_{new}$ | 1 | 2 | 0 | 1 | 0 | 0 | 0 | 0 | 0 | 1 | Process-4 | 1.0 | 0 |

**Table 19** Classifying New Test Process for Intrusion

|  | NearestNN | SimDist |
| --- | --- | --- |
| $P_{test}$ | Process-3 | Normal |
| $P_{new}$ | Process-4 | Abnormal |

## 5. CONCLUSION

Intrusion detection using text mining techniques has been recent research interest among researchers. The present approach of intrusion detection is based on applying clustering concept to perform dimensionality reduction. This dimensionally reduced process may then be used to perform classification and prediction. Through performing dimensionality reduction, we achieve space efficiency by reducing space complexity and also time efficiency by overcoming unnecessary computations which must be carried because of higher dimensions. In the present work, we use k-means and adopt the proposed distance measure to achieve clustering of processes. The tighter bound on distance value and the flexibility of Gaussian framework as helps us to carry incremental clustering of new incoming system processes. The present work of intrusion detection concentrates on static clustering of processes and is discussed using a case study. In future, we aim to perform the incremental clustering of the process to achieve efficient intrusion detection.